\documentclass[12pt,fleqn]{iopart}

\usepackage{iopams} 
\usepackage{harvard} 
\usepackage{graphicx}

\expandafter\let\csname equation*\endcsname\relax
\expandafter\let\csname endequation*\endcsname\relax 
\usepackage{amsmath} 

\usepackage{color}\definecolor{gray}{rgb}{0.5,0.5,0.5}

\def\fig#1{Fig.\,\ref{#1}}
\def\eq#1{Eq.\,(\ref{#1})}

\def\be#1{\begin{equation}\label{#1}}
\def\ee{\end{equation}}

\renewcommand{\vec}[1]{\mathbf #1}

\begin{document}

\title{Kicking electrons}

\author{Martin Gerlach, Sebastian W\"uster, and Jan M.\ Rost}

\address{Max Planck Institute for the Physics of Complex Systems\\
   N\"othnitzer Stra{\ss}e 38, 01187 Dresden, Germany}
\ead{rost@pks.mpg.de}
\begin{abstract}\noindent
The concept of dominant interaction hamiltonians is introduced to classical planar electron-atom scattering.
Each trajectory is governed in different time intervals by two variants of a separable approximate hamiltonian. Switching between them results in exchange of energy  between the two electrons. A second mechanism condenses the electron-electron interaction to instants in time and leads to an exchange of energy and angular momentum among the two electrons in form of kicks.  We calculate  the approximate and full classical deflection functions and show that the latter can be interpreted in terms of the switching sequences of the approximate one.
Finally, we demonstrate that the quantum results agree better with the approximate classical dynamical results than with the full ones.

\end{abstract}

\maketitle

\section{Introduction}
Approximations and simplifications are the key to physical understanding and therefore a 
large variety of approximations of different nature exist to date.
A physical system can  be described often by a Hamiltonian, the latter is  in most cases not separable permitting only numerical solutions. To obtain them becomes unrealistically time consuming already for a small number of degrees of freedom. Moreover, even if numerically tractable, there is often little insight gained from the numerical solutions. The same is partially true for the well known systematic perturbation theories.

Our concept of dominant interaction hamiltonians (DIH) aims at the formulation of an approximation which
can be solved analytically or numerically with ease, and at the same time provides insight into the dynamics. 
 To this end we approximate the full hamiltonian $H$ in different ways, $H_j$, $j = 1,2,\ldots$, valid in different regions of  phase space  which are visited in the course of time by a classical trajectory of the approximate system. Clearly, the concept requires classical dynamics to start with since it determines the relevant Hamiltonian $H_j$ through its local dominance in phase space.  
 However, as we will demonstrate,
the (classical) DIH approach produces a qualitative interpretation of the quantum dynamics of the system in terms of characteristic hamiltonian sequences $H_iH_jH_k\ldots$ which are classically realized through trajectories in the scattering process. 

We apply the concept of DIH to planar electron-atom scattering, more specifically, electron scattering from a He$^{+}$ ion in the ground state, a problem with enough intrinsic complexity to appreciate the qualitative picture of the dynamics which the DIH approach supplies. On the other hand, planar scattering is simple enough so that it can be handled quantum mechanically with a reasonable effort which enables us to gauge the DIH concept against exact quantum results. 
Somewhat surprisingly, DIH provides even a better quantitative approximation to the quantum results than the  exact classical solution.

The present work is also a logical next step in developing further our DIH concept which we firstly have applied successfully to high harmonic generation in formulation with one degree of freedom \cite{zago+12a,zago+12}.
 
In the next section we introduce the DIH concept and describe its prerequisites followed by  the 
formulation of the specific dominant interaction hamiltonians for electron ion scattering, using
the far field separation for the interaction among the two electrons. 
Section three presents the full classical results and those obtained with  DIHs in comparison. In section four we interpret the full classical dynamics with  the classification of trajectories emerging from the DIH approach and show that this allows us to identify and characterize the prominent peak structures in the full classical scattering cross section. Section five presents a comparison of our classical results with quantum calculations and demonstrates how DIH can be used to understand and approximate them.  Section six concludes the paper with a summary.

\section{Dominant interaction hamiltonians}
\subsection{The concept}
Let $\{H_i\}_{i=1,...,N}$ be a collection of hamiltonians  which all approximate the true hamiltonian $H$ of a system in different (reasonable) ways. The hamiltonian $H_{j}$ is called dominant  over a set $\Gamma_{j}$ of phase space points which is defined by $\Gamma_{j}=\{\gamma| H_{j}(\gamma) = \max_{i=1,...,N}H_{i}(\gamma)\}$, where $\gamma = (p,q)$. Hence, the phase space is partitioned according to segments $\Gamma_{j}$ with different dominant hamiltonians.
We construct  trajectories within $\Gamma_{j}$ according to Hamilton's equations with the dominant hamiltonian $H_j$ as usual,
\be{em}
\dot x = \frac{\partial H_{j}}{\partial p}\,,\,\,\,\,\,\,\,\,
\dot p = -\frac{\partial H_{j}}{\partial p}\,.
\ee
If the trajectory $\gamma(t)$ reaches at some time $s_{i}$ the boundary between two segments, e.g.,
$\gamma(s_i)\in \Gamma_{1} \cap \gamma(s_{i})\in \Gamma_{2}$, then the hamiltonian is switched for 
$t> s_{i}$ from $H_{1}$ to $H_{2}$, a procedure, which is repeated at all space boundaries a trajectory crosses.

\begin{figure}[bth]
\begin{center}
\includegraphics[width=.5\textwidth]{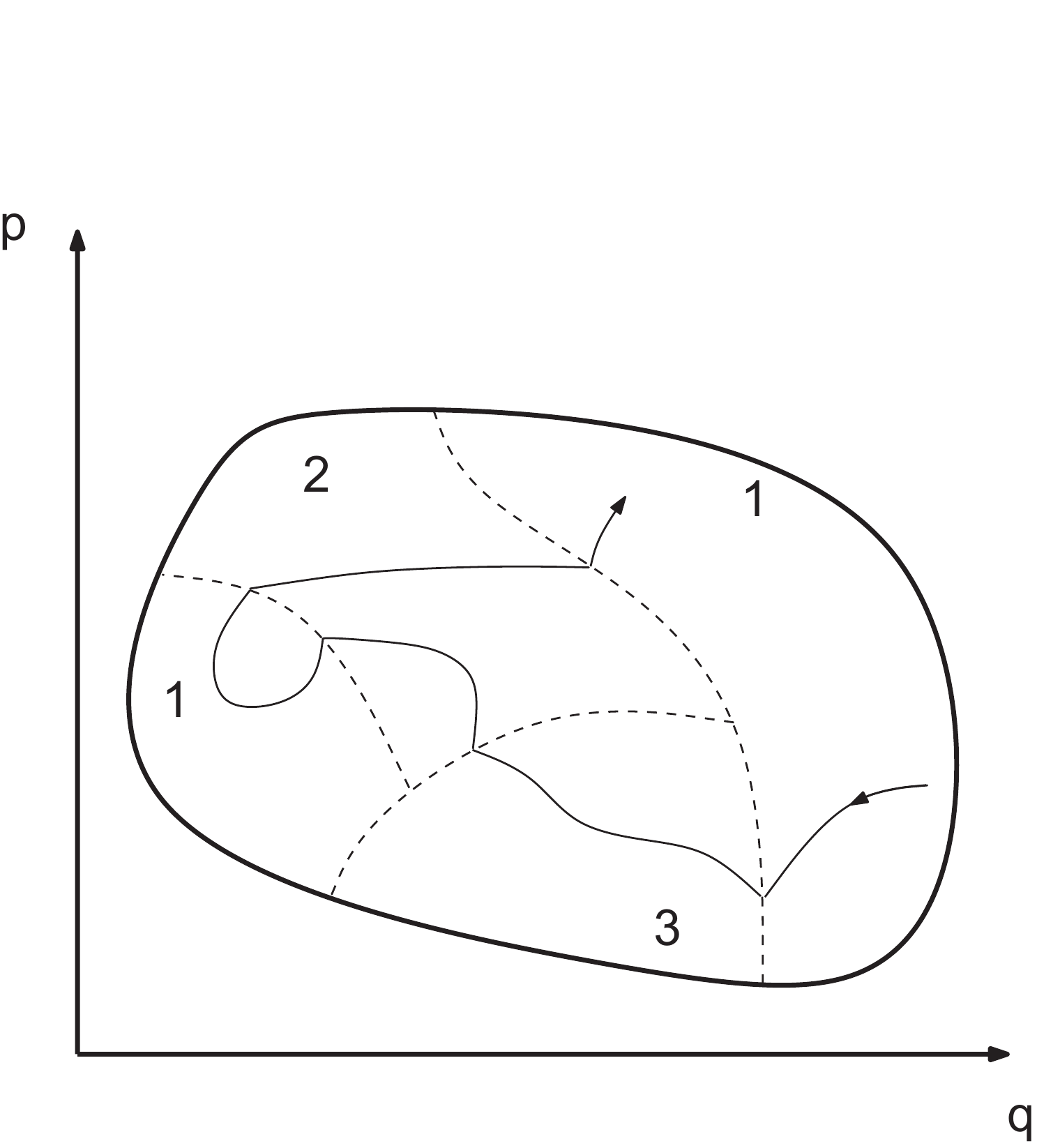}
\caption{Sketch of phase space partition through DIH. A  trajectory passing different DIHs in the sequence 132121 is also sketched.}
\label{fig:sketch}
\end{center}
\end{figure}

This construction leads to  a continuous but not necessarily differentiable trajectory, see the sketch in \fig{fig:sketch}. Each trajectory is characterized by the sequence of DIH, ($[132121]$ in \fig{fig:sketch}) which have been used to propagate it.

\subsection{Dominant interactions in three-body two-electron dynamics}
The abstract concept becomes much easier to grasp when applied to a specific example which will be the two electron problem with full hamiltonian
\be{ham2e}
\frac{\vec p_1^{2}}{2}+\frac{\vec p_2^{2}}{2}+ZV_{1}+ZV_{2}+V_{12}\,,\,\,\,\,\,\,\,\,\,\,\,\,V_{i}=
-\frac 1{r_{i}}\,,\,\,\,\,\,V_{12}= \frac 1{|\vec r_{1}-\vec r_{2}|}\,,
\ee
where the $\vec p_{i}$ are the two electron momenta and the $\vec r_{i}$ the position vectors pointing from the nucleus of charge $Z$ to the electron positions. 
What leads to energy and angular momentum exchange between the two electrons and renders the problem non-separable is the electron-electron interaction $|\vec r_{1}-\vec r_{2}|^{-1}$.

\subsubsection*{Far field switching}
In the {\it far field}, i.e., when both electrons are far away from each other, we can expand 
$|\vec r_{1}-\vec r_{2}|^{-1}$ over the electron which is further away from the origin, i.e., the nucleus. For $r_{2}\gg r_{1}$ this  gives in lowest order
$|\vec r_{1}-\vec r_{2}|^{-1} \approx 1/r_{2}$  which 
leads  to the separable hamiltonian
\be{farfieldH}
H_1= \frac{\vec p_{1}^{2}}2+\frac{\vec p_{2}^{2}}2-\frac Z{r_{1}} -\frac {(Z-1)}{r_{2}}\,.
\ee
The role of 
 $|\vec r_{1}-\vec r_{2}|^{-1}\approx 1/r_{2}$ 
in this case is simply to describe that the inner electron 1 screens the nucleus for the outer electron 2.
Of course,  $H_2$ also exists with the roles of electron 1 and 2 interchanged.
This approximation is also known as the  Temkin-Poet model \cite{te62,po78}, or restricted to radial coordinates $r_{i}$ only, also as the so called s-wave model \cite{hadr+93}. Here, $H_{1}$ and $H_{2}$ are DIH in their respective phase space domain $\Gamma_{i}$ and the switching (which consists in interchanging $\vec r_{1}\leftrightarrow \vec r_{2}$) takes place at $r_{1}=r_{2}$. 

\subsubsection*{Near field kicking}
\begin{figure}[tbh]
\begin{center}
\includegraphics[width=.5\textwidth]{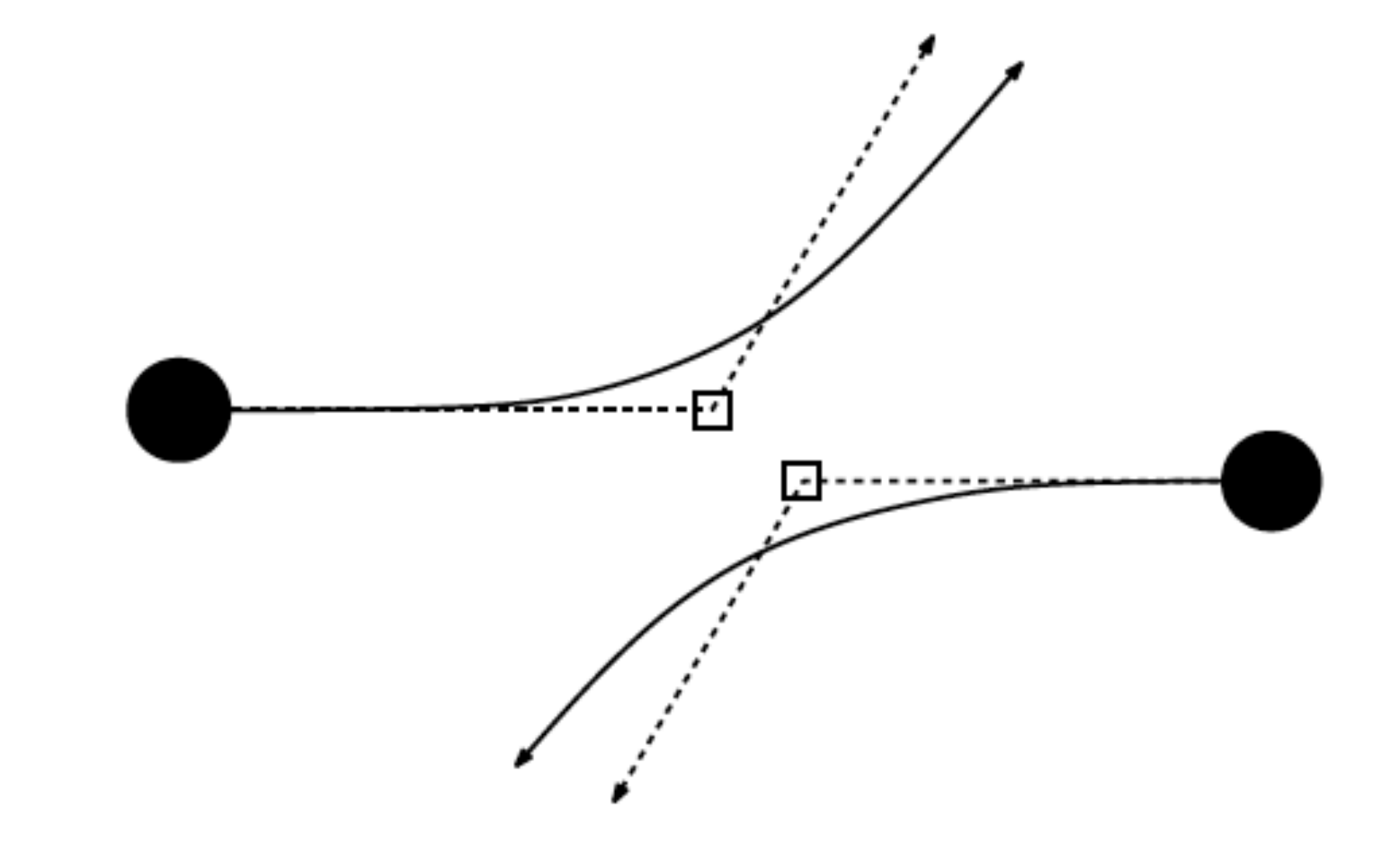}
\caption{Schematic illustration of the near field kick between two electrons under the condition
$ F = 1$ (\eq{Fratio}).}
\label{fig:repulsion}
\end{center}
\end{figure}

So far we have not discussed the 
{\it near field}, i.e., the situation that  the electron-electron interaction is  larger than the average electron-nuclear attraction,
\be{Fratio}
 F \equiv \left |\frac{2V_{12}}{Z V_{1}+Z V_{2}}\right| \ge 1\,.
\ee
A little thought reveals that if we would take  the corresponding separable hamiltonian
by neglecting $V_{i}$ in \eq{ham2e}
(with $\vec r = (\vec r_{1}+\vec r_{2})/2$ and $\vec R = \vec r_{2}-\vec r_{1}$)
\be{nearfieldH}
H _\mathrm{NF} = \vec P^{2}+ \frac {\vec p^{2}}{4}+ \frac 1R\,
\ee
as a dominant one for propagation, it would immediately counteract its dominance, since the purely repulsive interaction $1/R$ leads to increasing $R$ and therefore decreasing dominance of $H_\mathrm{NF}$.
Moreover, an energy preserving switching from one of the $H_{i}$ is difficult to achieve. It is much easier to assume that the effect of a purely repulsive DIH such as \eq{nearfieldH} can be concentrated to a single instant in time, where energy (and angular momentum) is exchanged among the electrons in a kick, without changing their positions and while respecting the constants of motion of the hamiltonian $H_\mathrm{NF}$. These circumstances provide sufficient conditions to uniquely define the kick
as sketched in \fig{fig:repulsion}.
The constants of motion $\vec{A}$, defined by a vanishing Poisson bracket $\{\vec A,H_\mathrm{NF}\}=0$, are given by $H_\mathrm{NF}$ itself, the total angular momentum $\vec L = \vec l_{r} + \vec l_{R}$ and the linear center-mass-momentum $\vec p$.   Since the kick is local 
at fixed distances $\vec r$ and $\vec R$, we have in addition that $\vec P^{2}=$const. as well as   $\vec l_{R}=$const.
From the last two conditions, one can construct a  transformation matrix $K_\mathrm{NF}$.
  Since $\vec L$ is conserved, the motion takes place in a plane where we take the interlectronic vector $\vec R = (X,Y)$ at the kick with $\tan\alpha = Y/X$.
Then the matrix $K_\mathrm{NF}$, which  transforms the vector $\vec P = (P_{x},P_{y})^{\dagger}$ before the kick into $\vec P' = K_\mathrm{NF}\vec P$ after the kick, can be parameterized with $\alpha$ as 
\be{trafoNF}
K_\mathrm{NF}  = \left( \begin{array}{rr}
-\cos2\alpha & -\sin 2\alpha  \\
-\sin 2\alpha & \cos2\alpha 
 \end{array} \right)
 =
 \left( \begin{array}{rr}
-1 & 0  \\
0 & 1
 \end{array} \right)
 \left( \begin{array}{rr}
\cos2\alpha & -\sin (-2\alpha)  \\
\sin (-2\alpha) & \cos2\alpha 
 \end{array} \right)\,.
\ee
Clearly, $|\det K_\mathrm{NF}| = 1$, since the modulus of the momentum is conserved, $P' = P$. The kick can be thought of as a rotation of the momentum vector by the angle $-2\alpha$ followed by an inversion of the $X-$component, as the product form in 
\eq{trafoNF} reveals. If, e.g., $\alpha = 0$, we have $\vec R = R\hat x$ such  that the force $-\nabla H_\mathrm{NF}$ leading to the kick acts in the direction of $\hat x$. Consequently, we get with \eq{trafoNF} in this case $P'_{x} = - P_{x}$ and
$P'_{y}=P_{y}$.

Taking into account the near field interaction in form of kicks completes our DIH formulation of two electron collision dynamics which uses for dynamical propagation exclusively the separable hamiltonian $H_{1}$ \eq{farfieldH} and its
counterpart $H_{2}$. The  conditions for switching and kicks and their consequences are summarized in table 1.

\begin{table}
\caption{Switching conditions of DIH for the two-electron problem with primed (unprimed) quantities indicating variables after (before) the switch; the function $F$ is defined in \eq{Fratio}. Note that total energy $E = E_1+E_2$ and total angular momentum $L = l_1+l_2$ are conserved.}
\label{tab:switch}
\vspace*{3mm}
\begin{tabular}{rcc}
\hline
event & '1' & '2'\\
condition & $F=1$ & $r_1=r_2$ \\
action & $\vec p'_{2}-\vec p'_{1}= K_\mathrm{NF}(\vec p_{2}-\vec p_{1})$ & $r_1\leftrightarrow r_2$\\
effect & $\Delta l_i\neq 0, \Delta E_i\neq 0$ & $\Delta E_i\neq 0$\\
\hline
\end{tabular}
\end{table}

\section{Planar classical electron-ion scattering}
\subsection{Hamiltonian}
For the practical implementation we restrict ourselves to total angular momentum $L=0$  which reduces
the degrees of freedom to the two electron-nucleus distances $r_{1},r_{2}$ and the angle $\theta$ 
between the vectors $\vec r_{i}$. The conjugate momentum $p_{\theta}$ can be viewed as the angular momentum of an individual electron $\vec l_{1}=p_{\theta}$, where $\vec l_{2}= -p_{\theta}$ to ensure
$\vec L = \vec l_{1}+\vec l_{2}\equiv 0$. The DIH hamiltonian corresponding to \eq{farfieldH} reads
\be{DIH}
H_{1}= \frac{p_{1}^{2}}2+\frac{p_{2}^{2}}2 +\frac{p_{\theta}^{2}}2\left(\frac 1{r_{1}}+\frac1{r_{2}}\right)-\frac Z{r_{1}}-\frac{(Z-1)}{r_{2}}\,.
\ee
\subsection{Initial conditions and the deflection function}
We assume electron 1 to be bound in the ionic ground state of He$^{+}$ with energy $E_{1}=-2$au,
and $l_{1}=0$ while electron 2 is the projectile starting with  energy $E_{2}$ (we use atomic units if not otherwise stated). 
 
\begin{figure}[bth]
\begin{center}
\includegraphics[width=\textwidth]{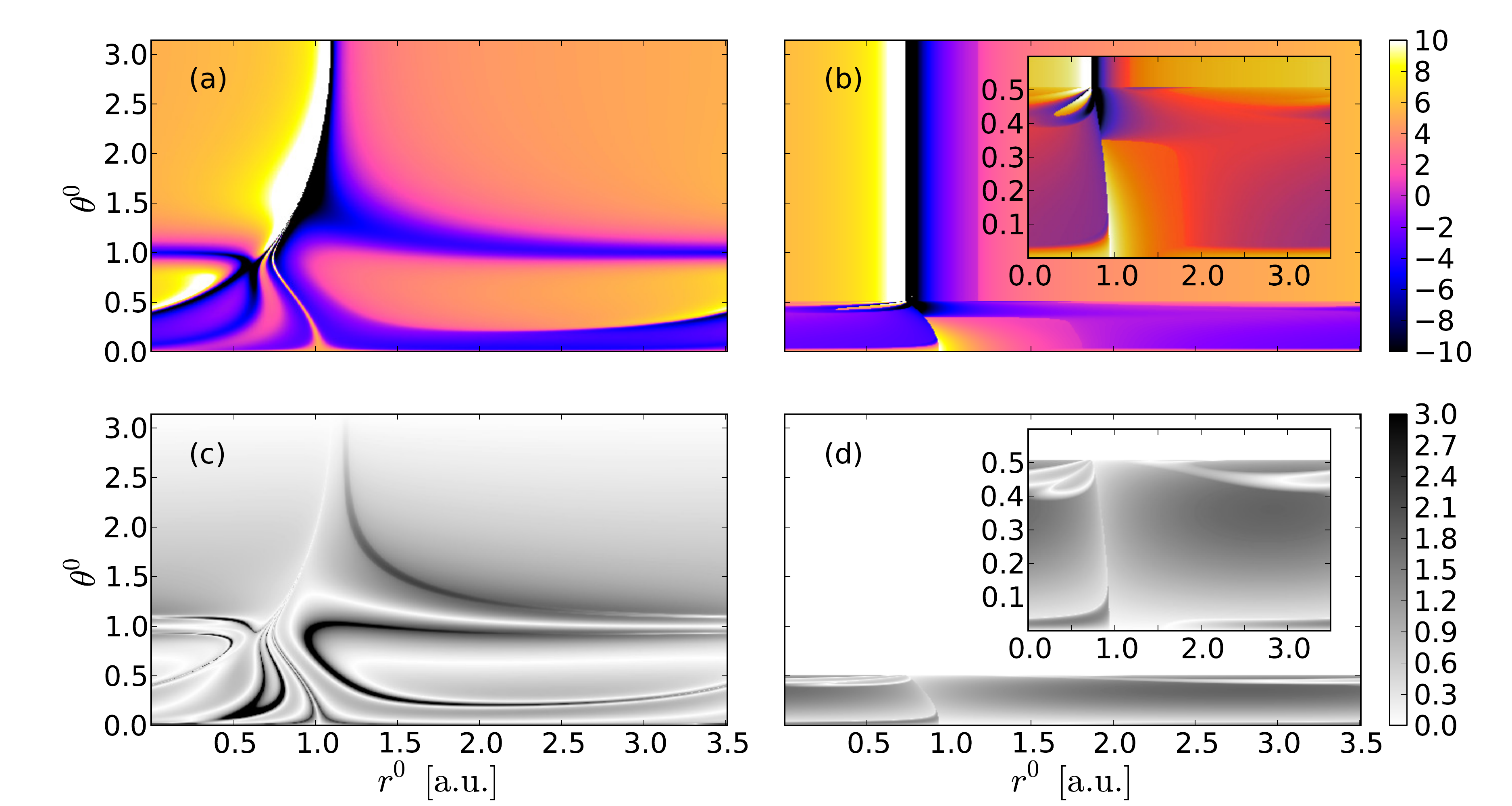}
\caption{Deflection functions for a collision with total energy $E = 0.5$ and initial
energy of the target electron of $E_{1}= - 2$. Shown is 
 $\epsilon^{*}(\theta^{0},r^{0})$ where
$\epsilon^{*}=E_{2}/E$ is the relative projectile electron energy  after the collision, (a) exact, (b) DIH dynamics and the final angular momentum $l^*(\theta^{0},r^{0})$ of the projectile electron (c) exact, (d) DIH dynamics.}
\label{fig:defl}
\end{center}
\end{figure}

For each classical trajectory we need to specify  6 initial conditions $(p_{1}^{0},r_{1}^{0},p_{2}^{0},r_{2}^{0},p_{\theta}^{0},\theta^{0})$. We let the trajectory of the bound electron always start at its outer turning point, i.e. $p_{1}^{0}=0,r_{1}^{0}=1$au. The projectile starts with momentum $p_{2}^{0}=-[2(E_{2}+1/r_{2}^{0})]^{1/2}$ where $r_{2}^{0}=1000$au$ + r^{0}$. Finally, $p_{\theta}^{0}=0$ (because $l_{1}=0$). Overall, this leaves two free variables $(\theta^{0},r^{0})$. Then, any probability to find a certain value $a$ for  the variable $A$ after  scattering can be formulated in terms of deflection functions $a^{*}(\theta^{0},r^{0}) \equiv \lim_{t\to\infty}A(t,\theta^{0},r^{0})$ \cite{ro98},
\be{defl}
\frac{dP}{da} = \frac{1}{\Delta\theta\Delta r}\int_{0}^{\Delta\theta} d\theta^{0}\int_{0}^{\Delta r}dr^{0}\delta(a - a^{*}(\theta^{0},r^{0}))\,,
\ee
where $\Delta\theta = \pi$ and $\Delta r = (E_{2}/8)^{1/2}\pi$ are the ranges of the initial variables. Therefore,
the important dynamical objects are the deflection functions.
They are shown in \fig{fig:defl} for the final energy $\epsilon$ and  the final angular momentum $l$ of the projectile.  Note that the deflection functions are periodic in $r^{0}$, since after the interval $\Delta r$ which corresponds to the distance the projectile travels during one period ($T=\pi/4$) of motion of the bound electron, the deflection function must repeat itself. Although the deflection functions seem to be quite different,  a closer look reveals that  full and DIH dynamics lead  to similar structural details for $\epsilon^{*}(\theta^{0},r^{0})$ but with different quantitative weights. Overall, the DIH structures appear to be concentrated within a much smaller range of initial values ${\theta^{0},r^{0}}$. In contrast, the DIH deflection function for the 
angular momentum $l^{*}(\theta^{0},r^{0})$ differs qualitatively since there is no change of the initial 
value $l=0$ for angles $\theta^{0} > \theta_{c}$. The reason is that angular momentum changing kicks according to the criterion \eq{Fratio} can only occur for $\theta < \arccos[1-1/(2Z^{2})] \approx 1/2$ 
for $Z=2$. Of course, \eq{Fratio} for the kicks can be modified  to increase the range of $\theta^{0}$ which can be changed during DIH dynamics. However, this leads to unphysically strong exchange of energy among the two electrons (recall that the effect of the kick is an exchange $\vec p_{1}\leftrightarrow \vec p_{2}$).

\subsection{Electron energy spectrum and angular momentum distribution}
\begin{figure}[bth]
\begin{center}
\includegraphics[width=.49\textwidth]{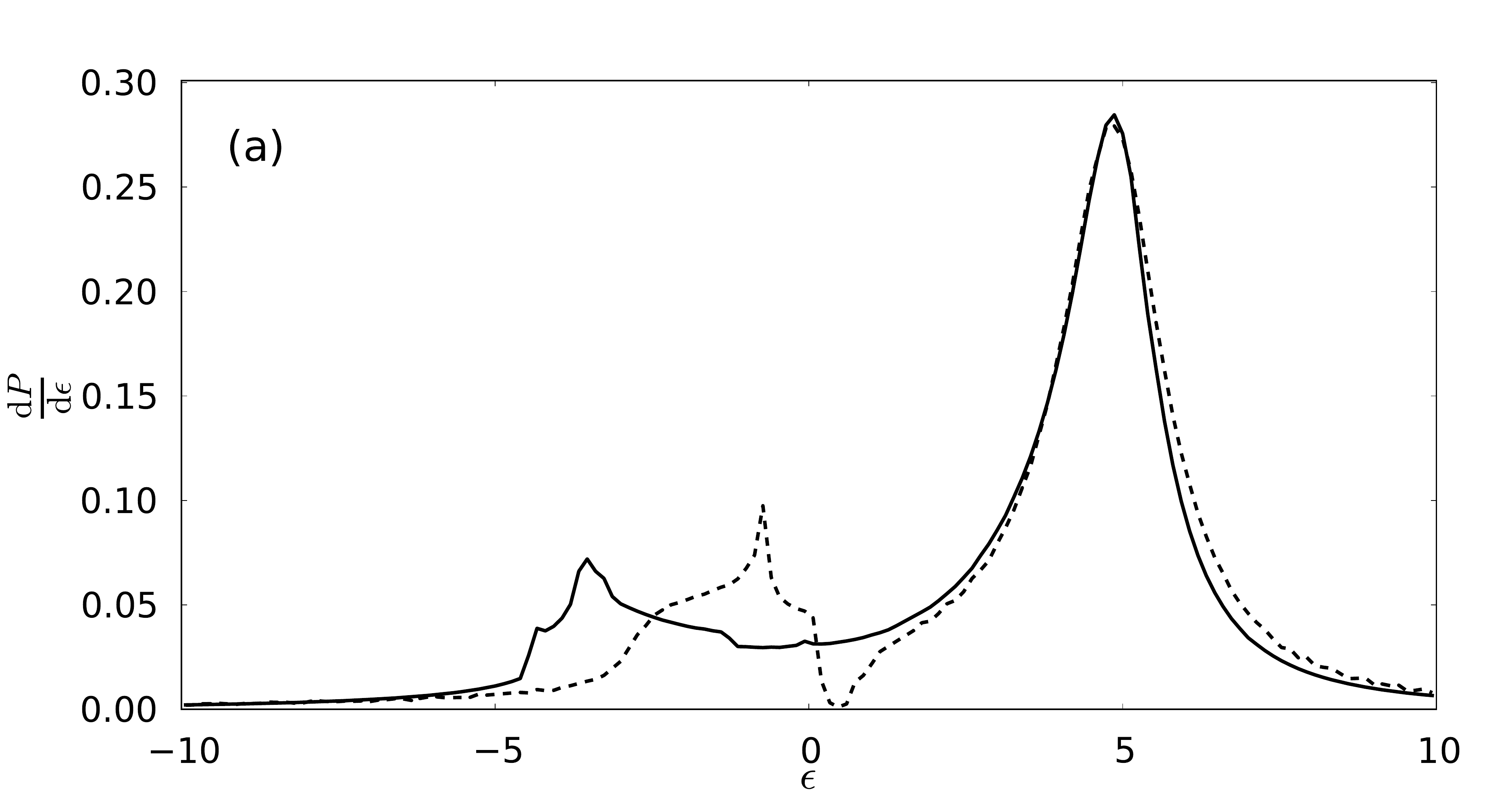}
\includegraphics[width=.49\textwidth]{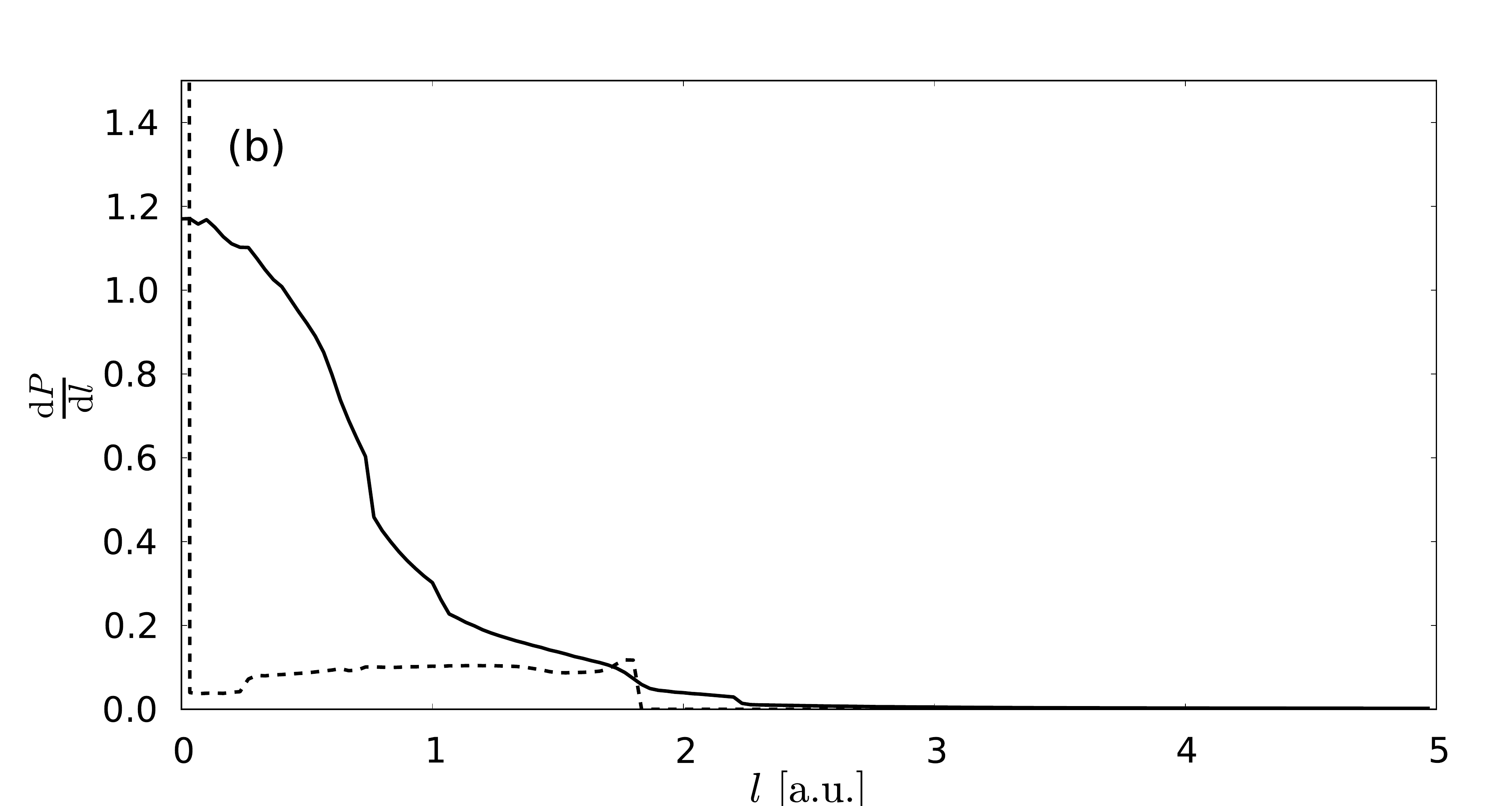}
\caption{Energy (a) and  angular momentum (b)   spectrum of the projectile electron under full (solid line) and DIH (dashed) dynamics for the the same collision as
in \fig{fig:defl}.}
\label{fig:specs}
\end{center}
\end{figure}

From the deflection functions one easily obtains the spectra for angular momentum and electron energy
by integrating  the respective deflection function over $\theta^{0},r^{0}$, see \eq{defl}. As one can see in \fig{fig:specs}a the
energy spectra of full and DIH dynamics agree qualitatively and even quantitatively for the 
large elastic scattering peak at $\epsilon= 5$ (which correpsonds to $E_{2} = 2.5 = E_{2}^{0}$).
The inelastic peak is of comparable magnitude but appears  shifted for the DIH dynamics.
In contrast, the angular momentum spectrum (\fig{fig:specs}b) differs considerably in  both 
approaches as is already apparent from the differences in the deflection function as discussed before.

However, as we will see later, this does not necessarily mean 
that the DIH dynamics gives poorer results compared to the {\it quantum} solutions than the full classical dynamics.  Before discussing the relation to the quantum results, we come, however, to the classification of trajectories and subsequently the entire dynamics which becomes possible through DIH.

\section{Classification of collision dynamics by sequences of dominant interaction hamiltonians}
The deflection functions in \fig{fig:defl} show a rich structure and the question arises if one can classify the different regions of initial conditions with characteristic 
properties of the trajectories starting from them. 
The DIH dynamics offers an obvious possibility, namely the sequence of DIH or, more precisely
 switches between DIH 1 and 2 (which we label as event '2') and application of kicks if $ F = 1$ (labeled subsequently as event '1'), where energy and angular momentum is exchanged among the two electrons.
 
\subsection{Typical DIH trajectories with switching events}
We first document the switching by presenting three typical cases  illustrated by the respective trajectories.

\begin{figure}[tbh]
\begin{center}
\includegraphics[width=.8\textwidth]{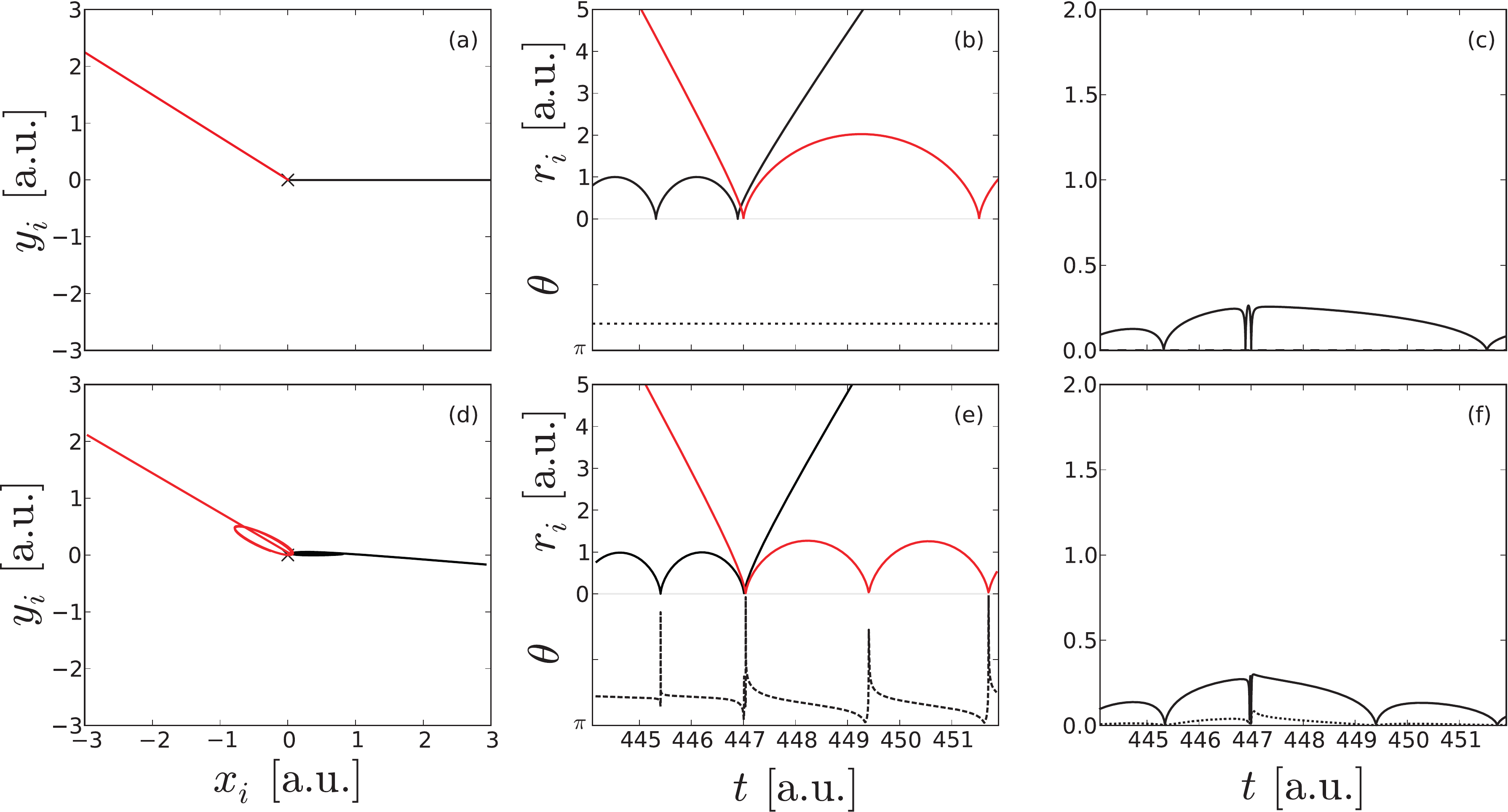}
\caption{Typical trajectories with single inelastic energy exchange by switching DIH hamiltonians, classified as event '2' in table \ref{tab:switch}, (a-c) DIH and (d-f) full classical dynamics. The general initial conditions for the collision are the same as in \fig{fig:defl}. The specific trajectory has the additional initial values $r_{2}^{0}= 1000+r^0$ with $(r^0,\theta^{0}) = (0.98,2.5)$ for (a-c) and $(r^0,\theta^{0}) = (1.12,2.5)$ for (d-f).  In the left panel the trajectories of the target (($x_{1},y_{1}$) - black) and projectile (($x_{2},y_{2}$) - red/light) electron are shown in space, the middle panel presents the radial evolution of the trajectories $r_{1}(t)$ and $r_{2}(t)$ (upper part) as well as $\theta(t)$ (lower part, dashed) in time, while the right panel records $F(t)$ (solid, \eq{Fratio}) and $\dot p_{\theta}(t)$ (dashed).}
\label{fig:event2}
\end{center}
\end{figure}

Excitation (i.e., an inelastic collision) can be achieved by event '2'. Excitation corresponds classically 
either to an increase of energy for the (still bound) target electron or to exchange of target and projectile electron, as it the case in \fig{fig:event2} (quantum mechanically, these two events cannot be distinguished). One can see in \fig{fig:event2}d that during the approach of the projectile small amounts of angular momentum are exchanged among the two electrons in the full dynamics while in the DIH approach the individual electron angular momenta remain zero throughout the trajectory (\fig{fig:event2}a). 
The (single) switching from $H_{1}$ to $H_{2}$ happens in this case very close to the nucleus and is difficult to see. 
Figs.~\ref{fig:event2}c,f confirm with the small changes in $F(t)$ and $\dot p_{\theta}$ that there is no angular momentum changing kick '1' involved. Consequently, the inter-electronic angle is basically constant for DIH dynamics (\fig{fig:event2}b) which holds on average also for the full dynamics. Only if the bound electron gets on its ellipse briefly on the other side of the nucleus, there is a spike at $\theta = 0$ (\fig{fig:event2}e). Overall, there is a good agreement between full and DIH dynamics.

\begin{figure}[bth]
\begin{center}
\includegraphics[width=.8\textwidth]{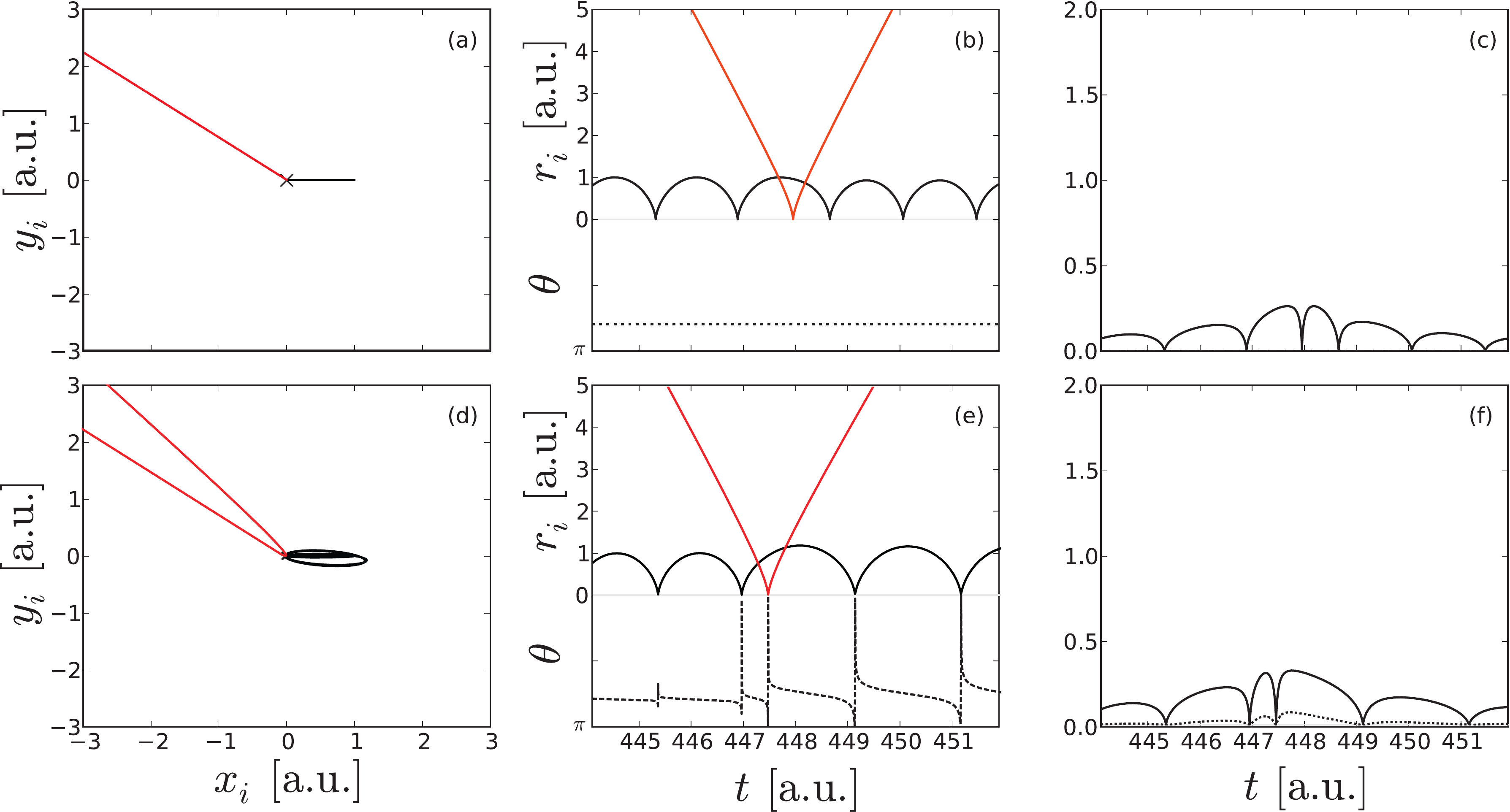}
\caption{Same as in \fig{fig:event2}, but for an inelastic collision with two switches, '22'.
The specific initial conditions are $(r^0,\theta^{0})  = (3.15,2.5)$ for (a-c) and $(2.10,2.5)$ for (d-f).}
\label{fig:event22}
\end{center}
\end{figure}

In \fig{fig:event22} switching  occurs twice constituting the event '22'. These trajectories correspond to a net excitation of the target electron which remains bound with a different energy than at the beginning of the collision.  One clearly sees the two switches where $r_{1}=r_{2}$, first $H_{1}\to H_{2}$ and then back to $H_{1}$. The dynamics in the angle $\theta$ is quite similar to the event '2', discussed before. Again, DIH and full dynamics are quite similar.

\begin{figure}[bth]
\begin{center}
\includegraphics[width=.8\textwidth]{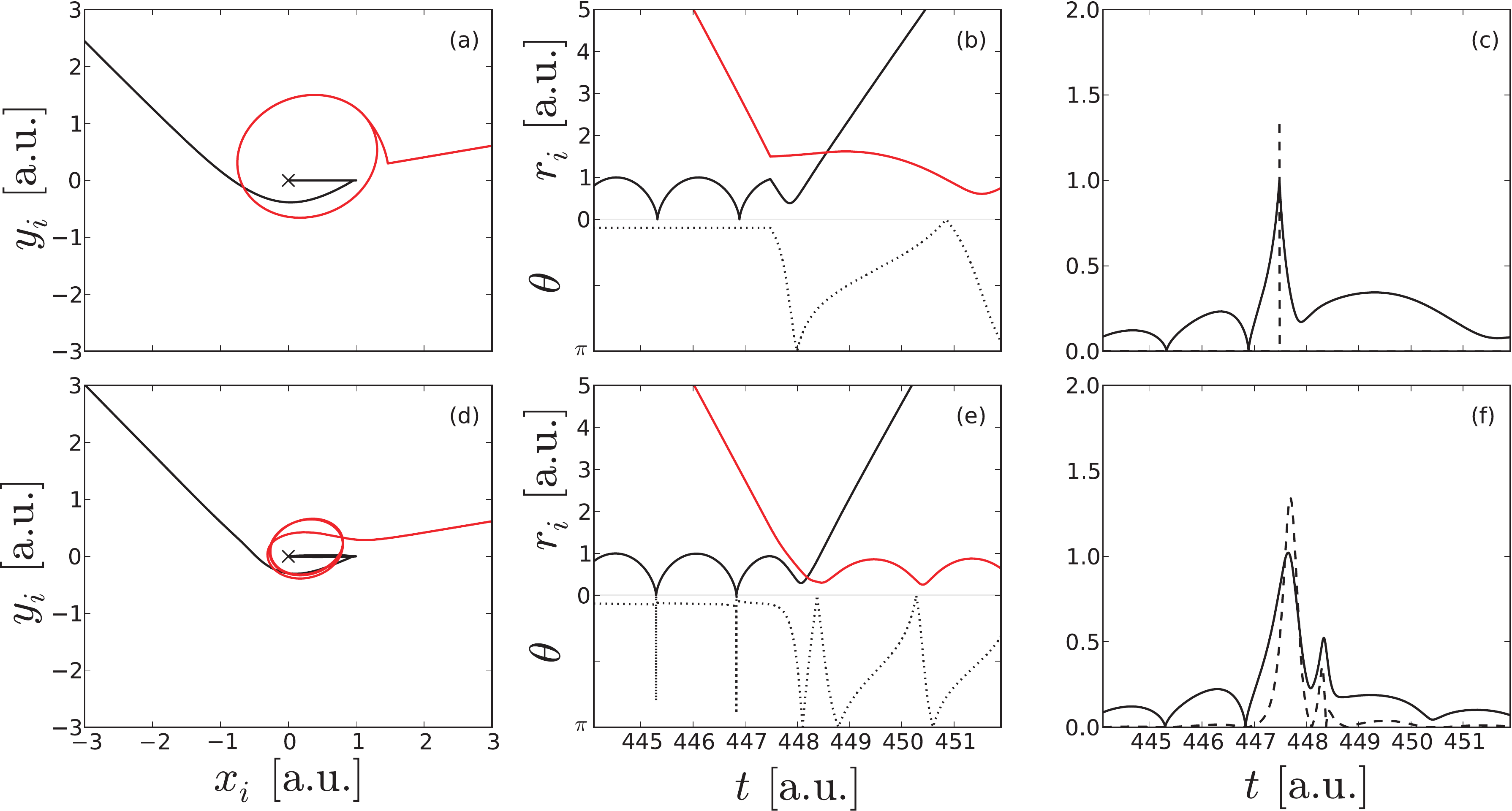}
\caption{Same as in \fig{fig:event2}, but for a collision sequence, '12'.
The specific initial conditions are $(r^0,\theta^{0})  = (3.15,0.2)$ for (a-c) and $(3.15,0.2)$ for (d-f).}
\label{fig:event12}
\end{center}
\end{figure}

Finally, we show in \fig{fig:event12} a collision with an event sequence '12'. Here, first through dominant electron-electron interaction ($F = 1$), quite a bit of single electron angular momentum is built up (see third panel). Indeed the criterion $F=1$ grasps the relatively sudden change in the angular momentum in the exact dynamics (dashed line in the third panel)  well. As one can see in the middle panel the angular momentum created induces of course motion in the angle $\theta$. Event '1' around $t=447.5$ is followed by an energy changing switch '2' around $t=448.5$. The exact and the DIH trajectory agree qualitatively, although not as good as in the two previous cases where no angular momentum dynamics was involved (only events of type '2'). This is consistent with our observation from the deflection function and the spectrum of the angular momentum and could be attributed to the restriction of events '1' in the DIH dynamics below a critical angle $\theta_{c}$.

\subsection{Classification of dynamics using switching sequences}
\begin{figure}[bth]
\begin{center}
\includegraphics[width=.8\textwidth]{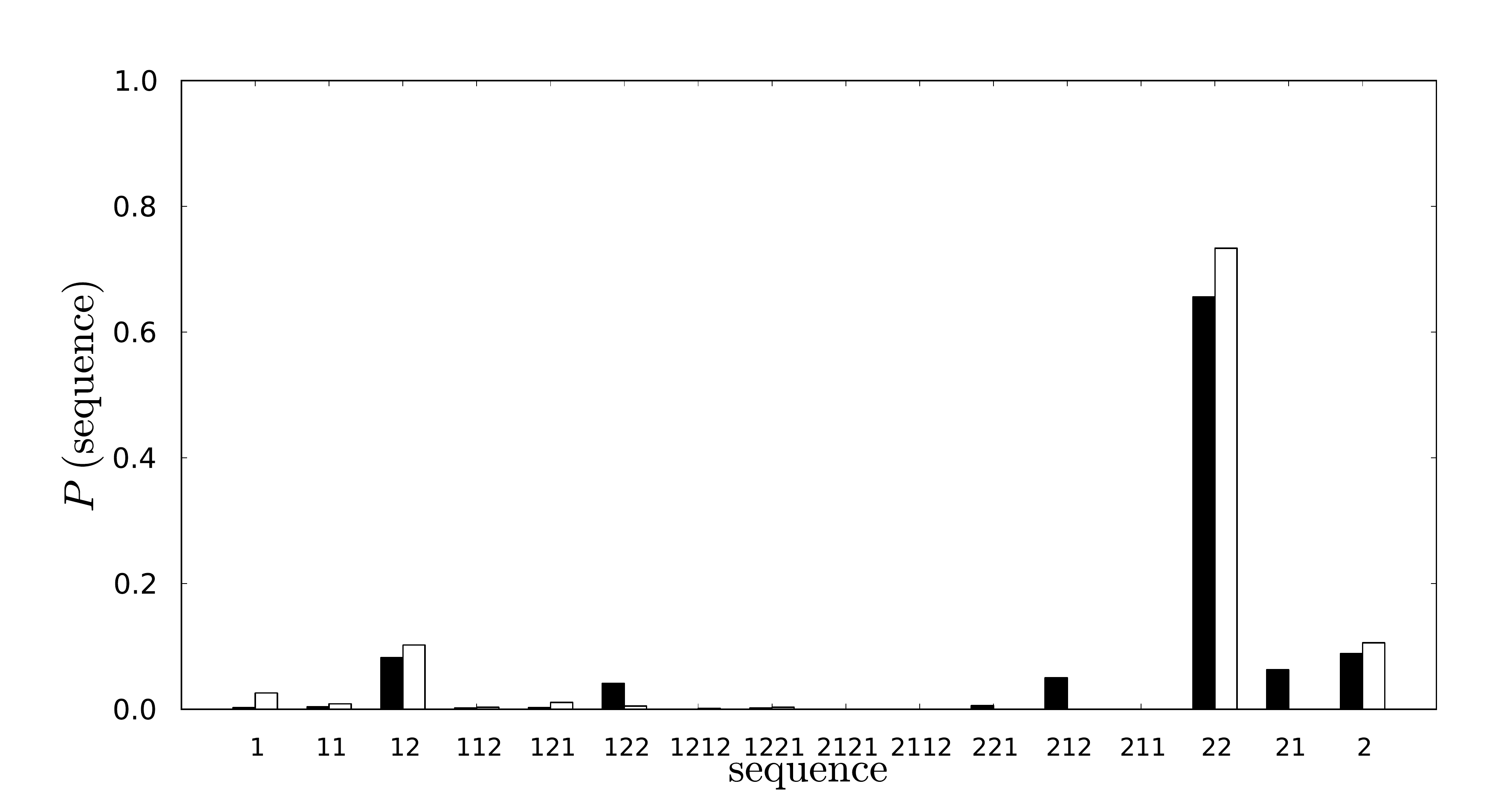}
\caption{Classfication of trajectories according to sequences of events '1' and '2' (see table \ref{tab:switch}) for  DIH (dark) and full (white) dynamics. }
\label{fig:sequences}
\end{center}
\end{figure}

\begin{figure}[bth]
\begin{center}
\includegraphics[width=.48\textwidth]{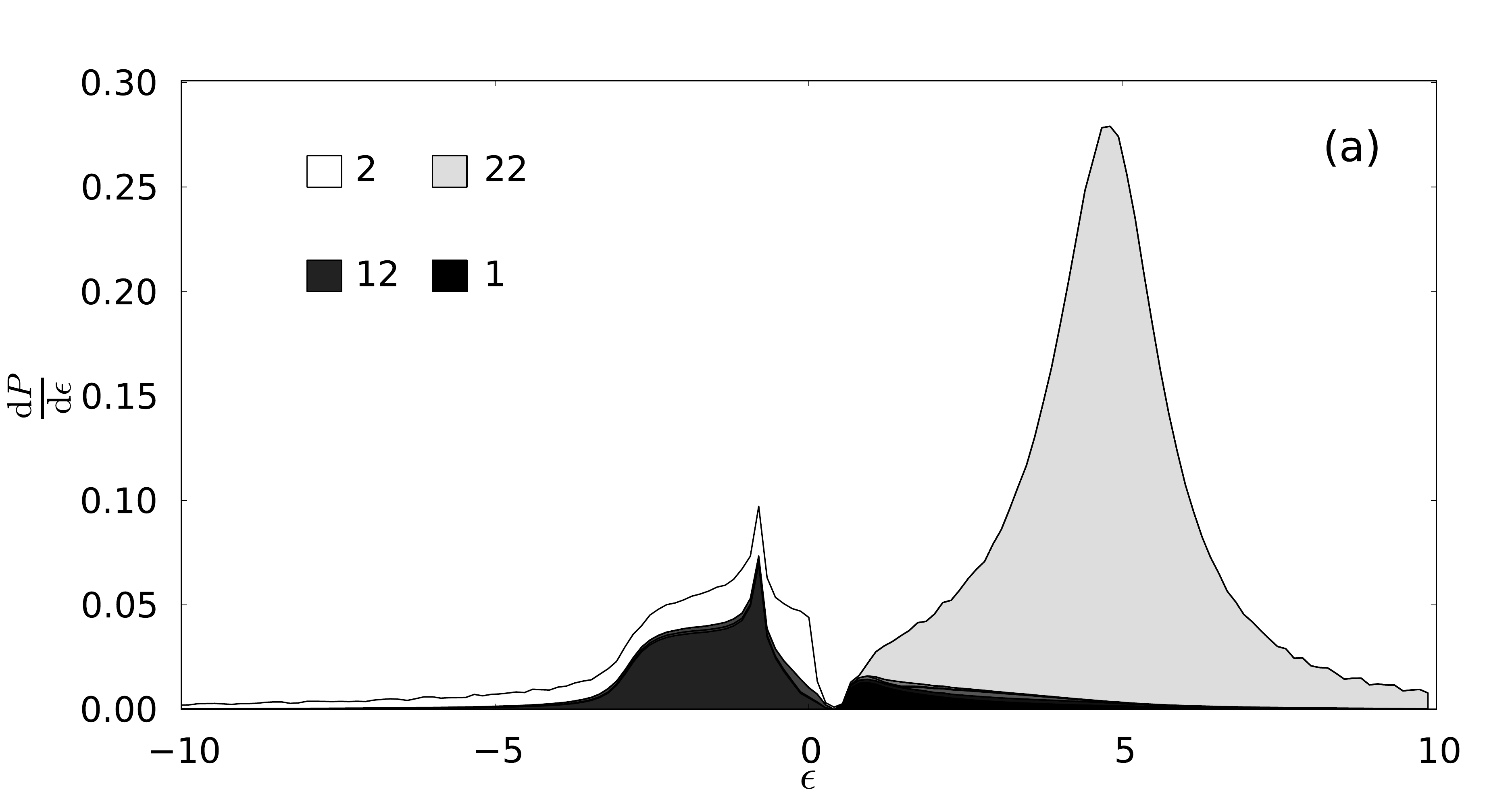}
\includegraphics[width=.48\textwidth]{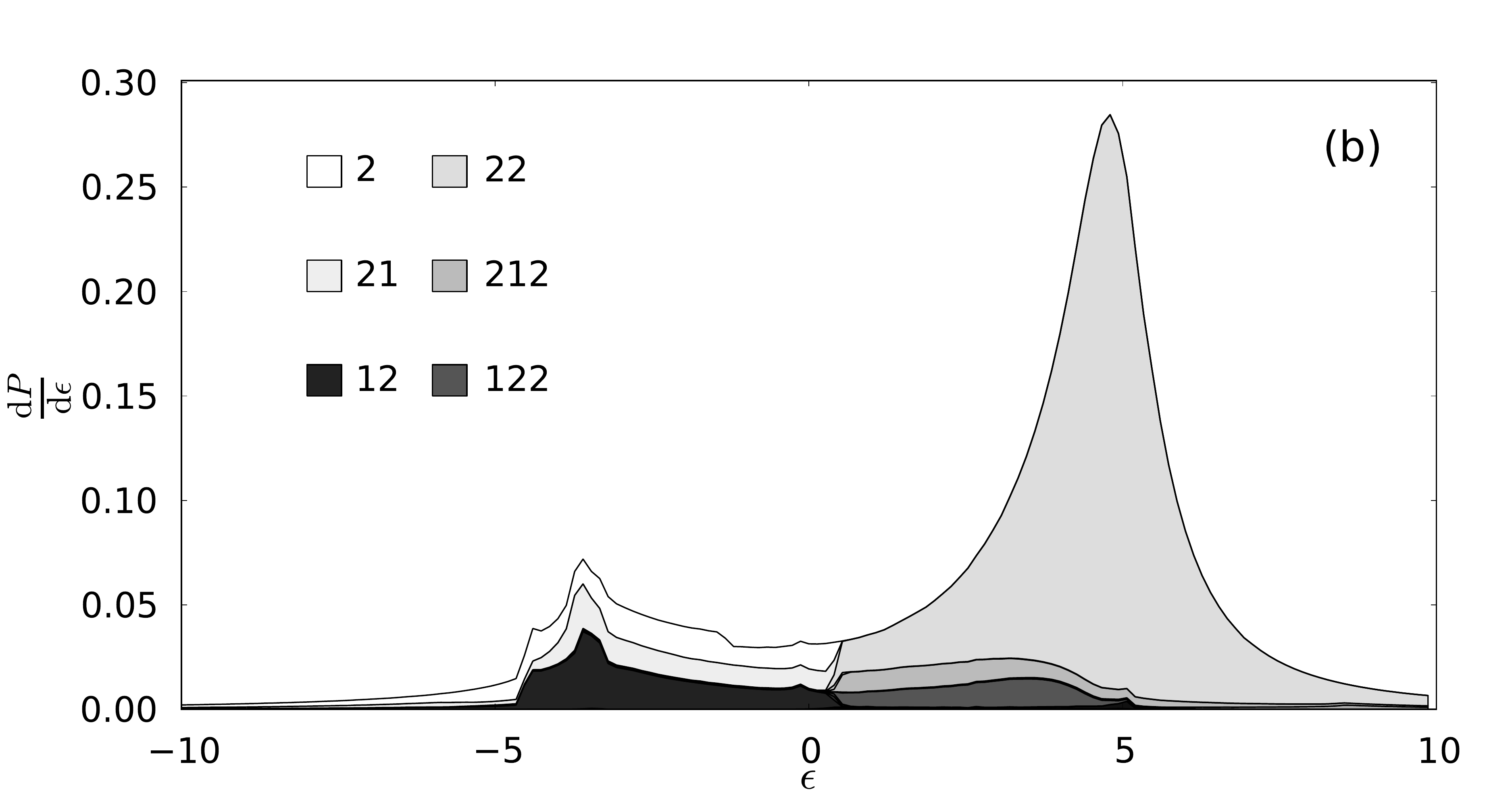}

\caption{Energy spectrum from \fig{fig:specs}a with contributions shaded according to their 
event sequences (see table \ref{tab:switch}), (a) for DIH dynamics (b) for full classical dynamics, for details, see text.}
\label{fig:espec-sequ}
\end{center}
\end{figure}

Overlooking all trajectories, the statistics of events
is quite similar (see \fig{fig:sequences}) with a clear dominance of '2', '22' and '12'.
Classifying contributions to the electron energy spectrum \fig{fig:specs}a according to the event sequences
in \fig{fig:espec-sequ} shows the meaning of the sequences: the elastic peak is clearly dominated by '22'
while the inelastic peak is built mostly from collisions of type '12'. Remarkably, this is not only true for the DIH dynamics, from which the classification originates, but also applies to the full dynamics.
This means, that even, if the DIH dynamics does not produce very accurate quantitative results, it can be used to generate a classification scheme which also applies to the full dymamics and is suitable to interprete and distinguish different mechanisms, such as elastic and inelastic collisions, etc.


\section{Comparison to quantum results}
\label{sec:quantum}
Our final task is to assess, how the full and approximate DIH classical dynamics performs in comparison to accurate quantum results.  To this end, we have developed a propagation scheme for the wave function on a grid in three dimensions which can handle the singular Coulomb interactions. It is described in the appendix.  Secondly,   the classical collision probabilities  have been symmetrized \cite{ro+95a} to obtain approximate singlet and triplet results.

\begin{figure}[bth]
\begin{center}
\includegraphics[width=.49\textwidth]{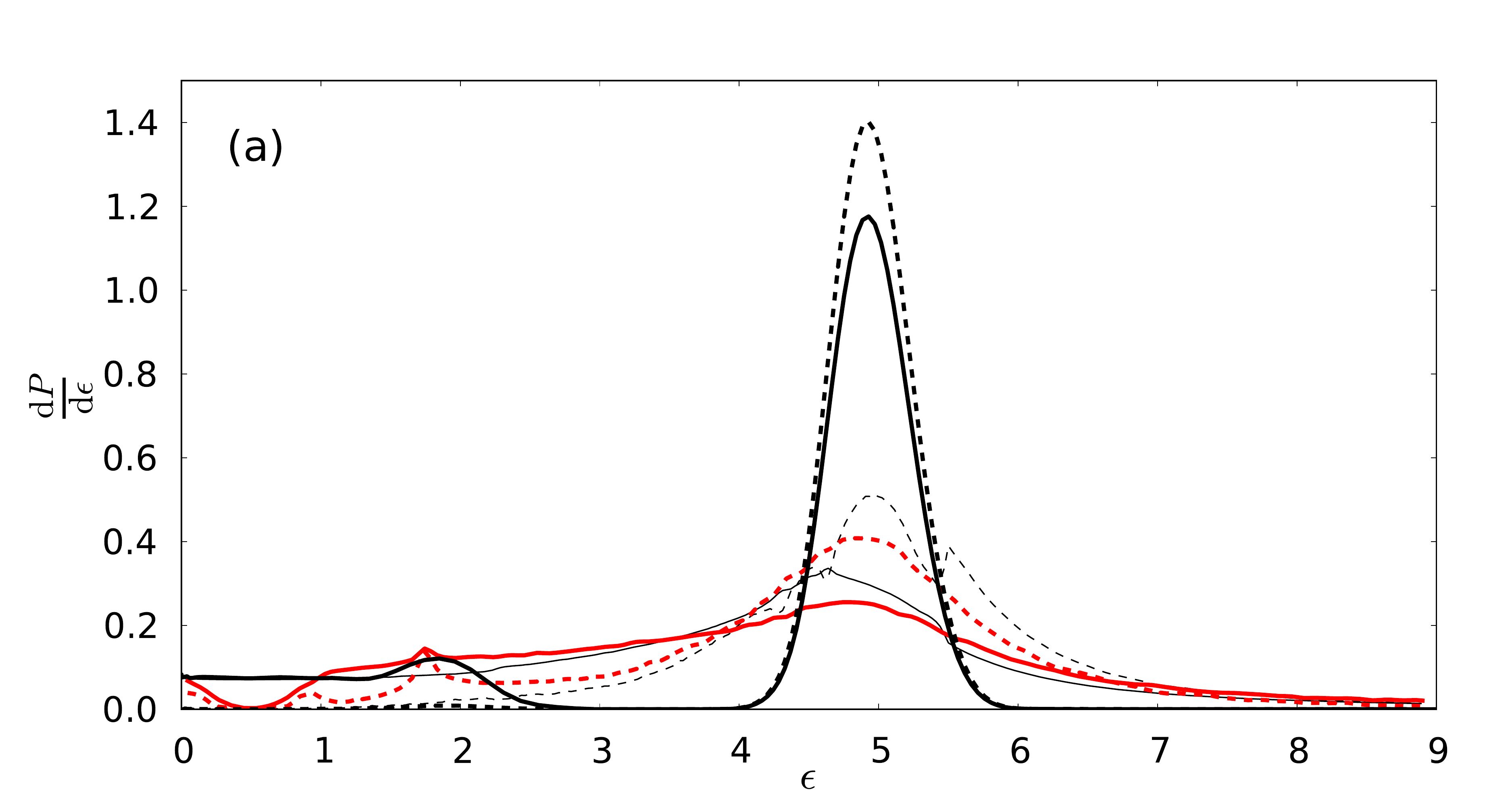}
\includegraphics[width=.49\textwidth]{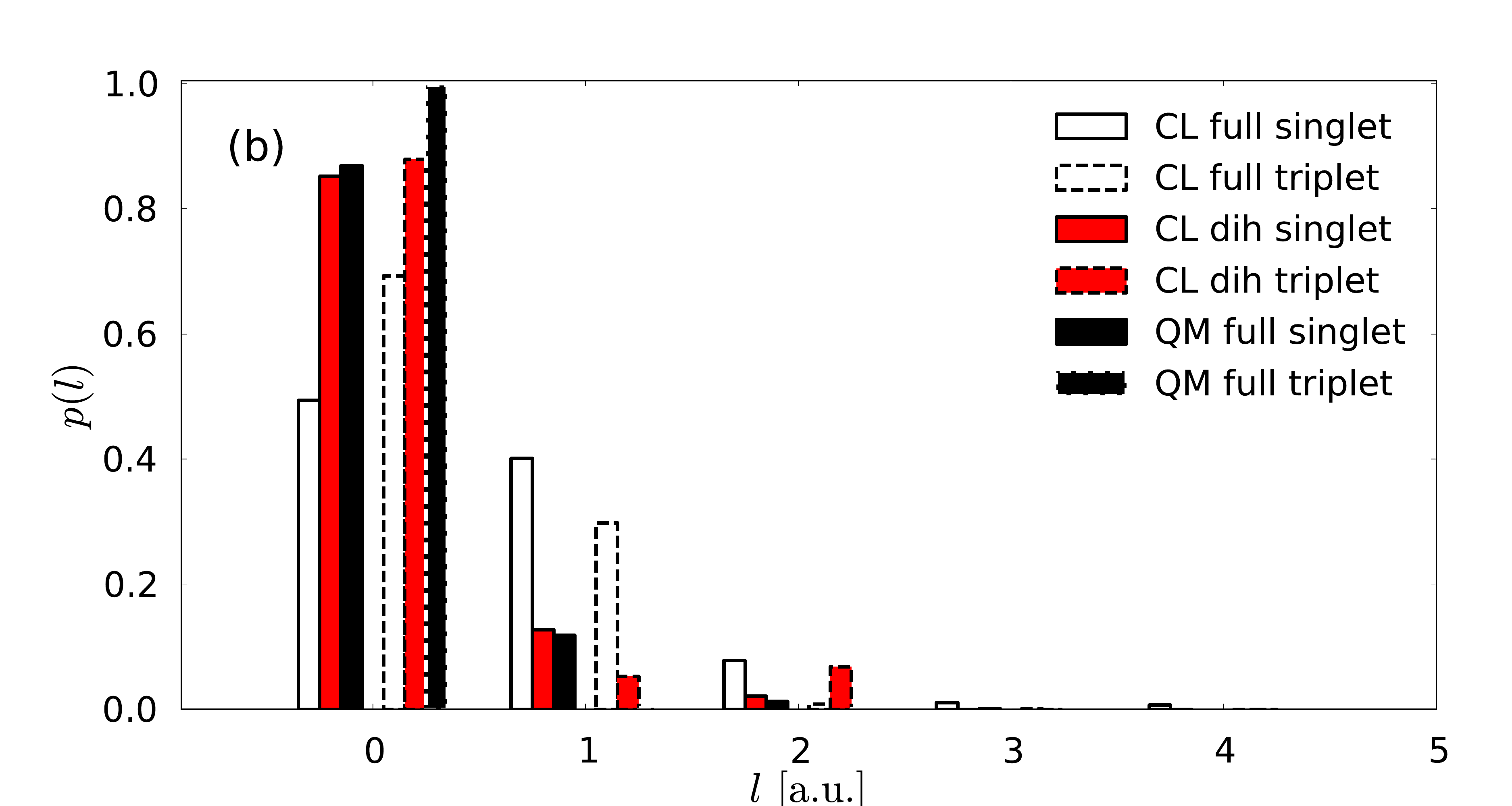}

\caption{Spectra of the projectile electron after the collision with parameter as in \fig{fig:defl} in singlet/triplet  symmetry (solid/dashed). The different curves provide the quantum result (thick/black), the classical full trajectory result (thin/white) and the classical DIH result (red). Part (a) gives the (continuous) energy spectrum, part (b) the binned distribution of final angular momentum of the projectile electron.}
\label{fig:spec-full}
\end{center}
\end{figure}

We have collided quantum electron wavepackets with the bound He$^+$ ion very similarly to the classical collisions process and obtain as a result the spectrum in \fig{fig:spec-full}. One sees that the elastic collision peak ($\epsilon = 5$) is roughly at the same position in  quantum and classical calculations (\fig{fig:spec-full}a). The quantum
peak is more concentrated about the elastic energy and therefore higher since only discrete excitation of the target electron is possible. Such excitation implies that the continuum electron needs to loose the excitation energy. The corresponding  peaks in the region $\epsilon \le  2$ for the singlet spectrum are not resolved due to the initial wave packet with its finite energy width but lead to a  smooth maximum in the quantum spectrum.  
Even higher excitation energies lead to lower final momentum for the projectile and in this semiclassical regime  quantum and classical spectra come together.

Similar considerations for the angular momentum spectrum require a discretization of the continuous classical angular momentum which can be done by binning \cite{lepe+78}. The comparison  shown in \fig{fig:spec-full}b reveals
that the symmetrized DIH result is in  better quantitative agreement with the quantum  spectrum, in particular for the triplet symmetry, than the full classical calculation. This may be attributed to the fact that the quantum triplet dynamics is less reactive than the singlet dynamics due to a symmetry enforced nodal line at $r_{1}=r_{2}$. Classically, this effect is resembled to a certain degree by the 
DIH dynamics compared to the full classical dynamics since in the former  ``reactivity'' is limited to the events '1' and '2', discrete in time.

\section{Summary}
We have introduced the concept of dominant interaction hamiltonians which approximates dynamics described by a complicated, non-separable classical hamiltonian with different simplified hamiltonians. Each of them is valid in a specific phase space volume where it dominates all other simplified hamiltonians formulated.
Applied to planar electron-ion scattering, we have demonstrated that the DIH approach provides a good approximation to the full classical dynamics. More importantly, and somewhat surprisingly, quantum results regarding differential spectra (energy and angular momentum of the projectile) agree better with the  DIH result than with the full classical dynamics. Whether this is accidental or systematic will have to be investigated in future studies. A second appealing aspect of the DIH concept is the qualitative picture it generates for the dynamics through the sequence of DIH hamiltonians passed by trajectories. We could show that prominent peaks in the quantum mechanical differential energy spectrum can be associated and therefore interpreted  with characteristic DIH sequences. This opens the way to classify and understand complicated dynamics through DIHs. For further quantitative improvement of DIH  the next natural step will be the formulation of a semiclassical extension of DIH.

\appendix\section{}
\subsection*{Numerical propagation of singular hamiltonians for three degrees of freedom}

%
%
%
In this section we give a detailed account of the propagation scheme used to numerically solve the time-dependent Schroedinger equation (TDSE) for the two-electron problem in section \ref{sec:quantum}.
Applying the infinite-nucleus approximation and restricting ourselves to the case of zero total angular momentum, the number of degrees of freedom reduces to three and the corresponding hamiltonian in coordinate representation is (cf. \eq{DIH})
\begin{equation}
  H = -\frac{1}{2}\frac{\partial^2}{\partial r_1^2}-\frac{1}{2}\frac{\partial^2}{\partial r_2^2}
     -\frac{1}{2}\left(\frac{1}{r_1^2}+\frac{1}{r_2^2}\right)\frac{1}{\sin\theta}\frac{\partial}{\partial \theta}\left(\sin \theta \frac{\partial}{\partial \theta}\right) + V\left(r_1,r_2,\theta\right),
\label{eq:HamFull}
\end{equation}
with the potential
\begin{equation}
 V(r_1,r_2,\theta)=-\frac{Z}{r_1}-\frac{Z}{r_2}+\frac{1}{\sqrt{r_1^2+r_2^2-2r_1r_2\cos\theta}}\,.
\label{eq:FullPotential}
\end{equation}
This system has been treated 
with a finite-difference method  in \cite{zhfe+94}, our approach presented here uses a spectral method.

\subsection*{Algorithm}
For the direct numerical integration of the TDSE we use the script language \textit{xmds} (www.xmds.org) 
which offers a variety of algorithms for solving partial differential equations.
As the wavefunction $\psi$ is represented on a discretized grid in the coordinates ($r_1,r_2,\theta$), the partial derivatives with respect to all three spacial variables are evaluated by means of fast-Fourier transform (FFT).
For the evolution in time we employ an explicit $8^{\mathrm{th}}/9^{\mathrm{th}}$ order Runge-Kutta method with an adaptive time step, enabling us to put an upper limit of $10^{-8}$ for the relative error per timestep.
Furthermore, xmds allows for an easy parallelization of the simulation.
A detailed description of the algorithm can be found in the documentation of xmds \cite{coce+08}.

\subsection*{Definition of the grid}
The crucial step is to define a suitable grid in the coordinates ($r_1,r_2,\theta$), thereby accounting for the singularities of the hamiltonian of Eq. \eqref{eq:HamFull} at $r_i=0$ and $\sin \theta =0$, the boundary conditions for the wavefunction, and the long-range character of the Coulomb interaction, which is especially important in electron-atom scattering.

In each coordinate $x$, where $x= r_1,r_2$ or $\theta$, the grid consists of $N_{x}$ points which are distributed equidistantly (due to FFT) over the interval $[x_{\min},x_{\max}]$. The positions follow from arranging the singularity just between two grid points, which allows a treatment of the full Coulomb potential.

Since the algorithm requires periodic boundary conditions, we define the wavefunction in each coordinate in an interval $[-L_x,L_x]$ according to 
\begin{subequations}
 \begin{align}
   \psi(-\theta)&=\psi(\theta),\,\,\,\theta>0,\\
   \psi(-r_{1,2})&=-\psi(r_{1,2}),\,\,\,r_{1,2}>0,
  \end{align}
\end{subequations}
where the second equation guarantees the additional boundary condition $\psi(r_{1,2}=0)=0$.
However, this implies that $\psi(r_{1,2} \rightarrow L_{r_{1,2}}) = 0$, so that $L_{r_{1,2}}$ has to be chosen large enough, especially in a scattering experiment.

In order to achieve a higher grid point density near the Coulomb singularity we employ a transformation of the radial coordinates according to \cite{faba+96}
\begin{equation}
  r_{1,2}=q_{1,2}-c_0\arctan(c_1 q_{1,2}),
\label{eq:TrafoGrid}
\end{equation}
with  coefficients $c_0$, and $c_1$, so that the radial grid (and thus the TDSE) has to be formulated with respect to the coordinates $q_{1,2}$.
The parameters for the grid in the coordinates ($r_1,r_2,\theta$) using Eq. \eqref{eq:TrafoGrid} are chosen as follows:
\begin{equation}
 N_{\theta}=16,\,\,\,L_{\theta}=\pi,\,\,\,N_{q_1}=N_{q_2}=256,\,\,\,L_{q_1}=L_{q_2}=125\,\mathrm{a.u.},\,\,\,c_0=18.0,\,\,\,c_1=0.05.
\label{eq:NumParam}
\end{equation}

\subsection*{Numerical tests}
As a test for the algorithm and the choice of our grid defined in Eq. \eqref{eq:NumParam}, we numerically calculate the spectrum $\sigma(E)$ for different hamiltonians by Fourier transform of the correlation function $c(t)$ \cite{ta07}:
\begin{equation}
 \sigma(E) \sim \int \mathrm{d}t\, e^{i E t} c(t),
\end{equation}
where $c(t)= \int \mathrm{d}x\, \psi(x,0)^* \psi(x,t)$ with the wavefunction $\psi(x,t)$ at time $t$.
\subsection*{Angular grid}
\begin{figure}[h]
 \begin{center}
   {\includegraphics[width=0.5\textwidth]{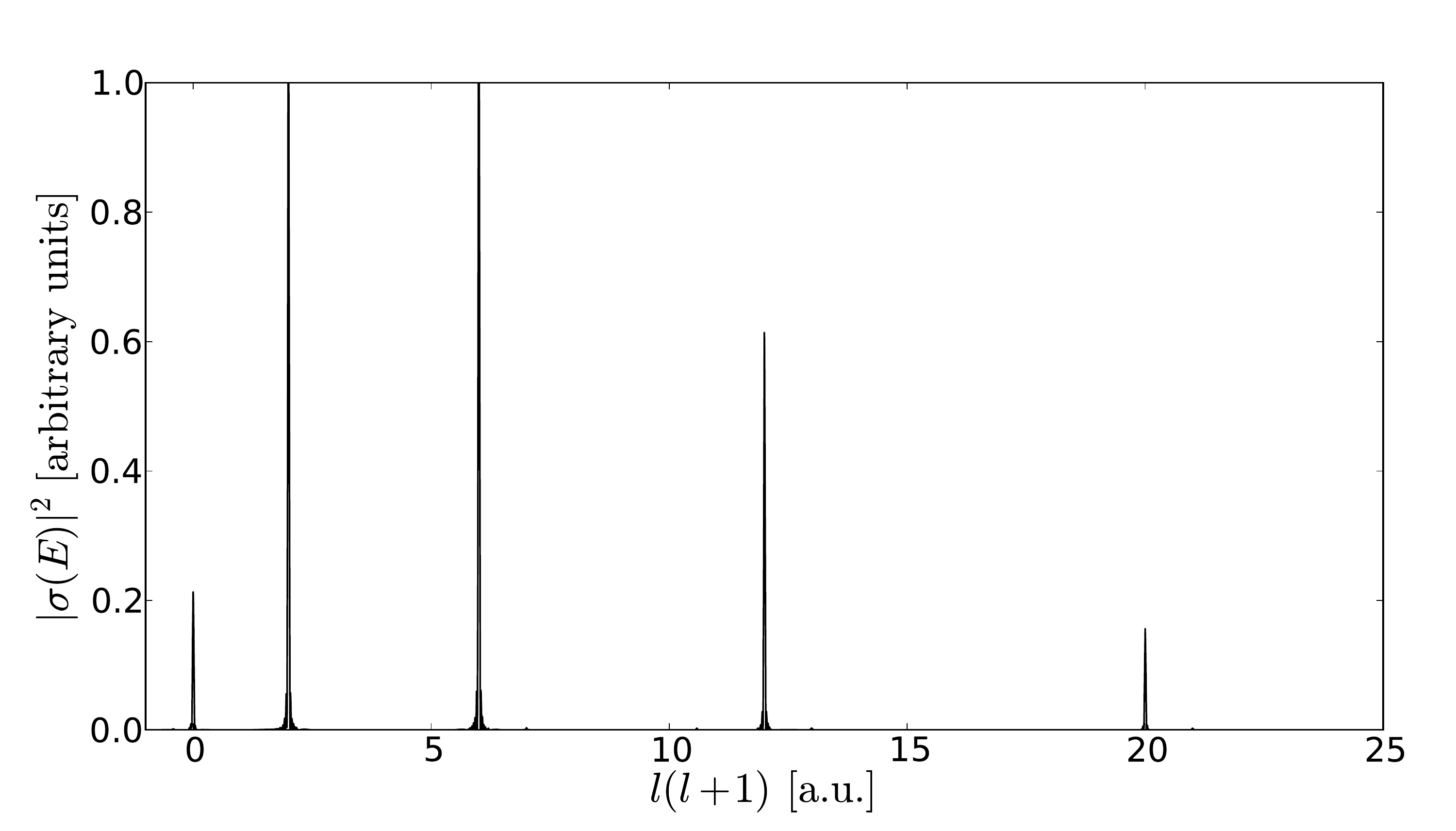}}
 \end{center}
  \caption[NumTestAngleSpec]{Spectrum $\sigma(E)$ of the hamiltonian Eq. \eqref{eq:HamAng} obtained through numerical integration of the TDSE for $t=200\,\mathrm{a.u.}$ with the initial wave packet of Eq. \eqref{eq:PsiInitAng} and $N_{\theta}=16$ grid points.}
  \label{fig:NumTestAngleSpec}
\end{figure}

For the evaluation of the angular grid we look at the angular part of the kinetic energy in Eq. \eqref{eq:HamFull}:
\begin{equation}
  H_{\theta}= -\frac{1}{\sin\theta}\frac{\partial}{\partial \theta}\left(\sin \theta \frac{\partial}{\partial \theta}\right),
\label{eq:HamAng}
\end{equation}
which resembles an angular momentum operator with eigenvalues $l(l+1)$.
Propagation of the initial wavepacket 
\begin{equation}
  \psi(\theta,0)= e^{-4(\theta - \pi)^2}
  \label{eq:PsiInitAng}
\end{equation}
for a time of $t=200\,\mathrm{a.u.}$ yields the spectrum shown in Fig.\ref{fig:NumTestAngleSpec}.

The peaks reproduce the five lowest eigenenergies of Eq. \eqref{eq:HamAng}, which means that the dynamics in the coordinate $\theta$ is well described at least for moderate angular momenta.
\subsection*{Radial grid}
\begin{figure}[h]
 \begin{center}
   {\includegraphics[width=0.5\textwidth]{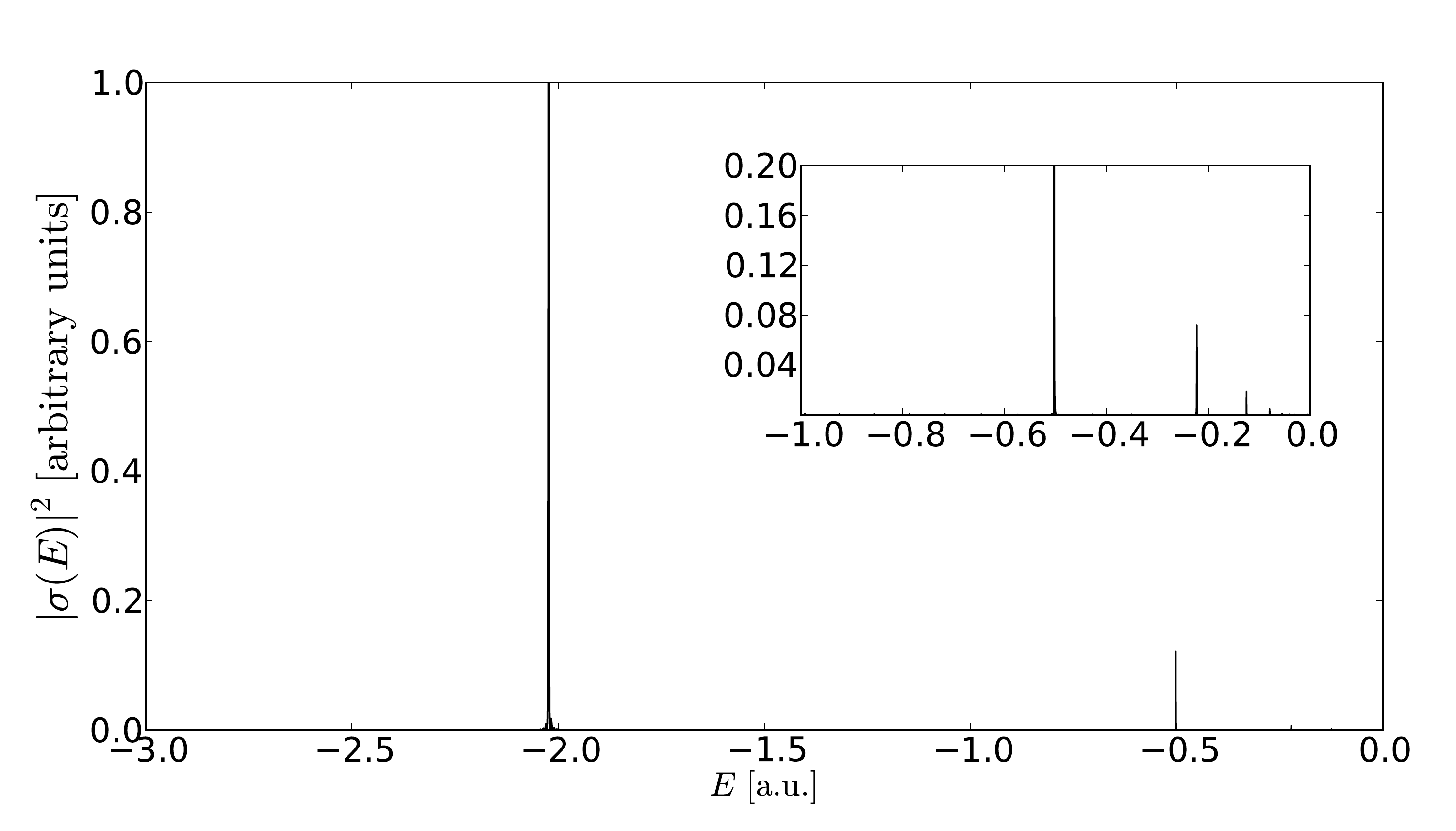}}
 \end{center}
  \caption[NumTest1DHSpec]{Spectrum $\sigma(E)$ of hamiltonian Eq. \eqref{eq:HamRad} obtained through numerical integration of the TDSE for $t=10000\,\mathrm{a.u.}$ with initial wave packet of Eq. \eqref{eq:PsiInitRad}. The radial grid was transformed according to Eq. \eqref{eq:TrafoGrid} with parameters from Eq. \eqref{eq:NumParam}.}
  \label{fig:NumTest1DHSpec}
\end{figure}

For the electron-He$^{+}$ scattering in section \ref{sec:quantum}, the target electron is prepared in the ground state while the influence of the projectile electron is neglected initially.
Therefore, we examine the radial component of the hydrogen problem, described by the hamiltonian
\begin{equation}
 H_r=-\frac 12\frac{\partial^2}{\partial r^2} -\frac{2}{r},
\label{eq:HamRad}
\end{equation}
to check if the radial grid can account for the bound motion near the nucleus.
Propagation of the initial wavepacket
\begin{equation}
 \psi(r,0)=re^{-r^2}
\label{eq:PsiInitRad}
\end{equation}
for a time of $t=10000\,\mathrm{a.u.}$ yields the spectrum shown in Fig.~\ref{fig:NumTest1DHSpec}.
There are distinct peaks  at energies $E=-2.02$, $-0.502$, $-0.224$, $-0.128$, $-0.081\,\mathrm{a.u.}$, which correspond to the five lowest eigenenergies $-2/n^{2}$ of Eq. \eqref{eq:HamRad} with a relative error of $<10^{-2}$.
The difference in the amplitude of the peaks is due to the smaller overlap of the initial wavepacket with higher excited eigenstates, which accounts for the fact that only four peaks are visible in Fig. \ref{fig:NumTest1DHSpec}.
This result shows that the radial grid enables us to describe the dynamics of the bound electron to a sufficient degree.

\subsection*{Full grid}
\begin{figure}[h]
 \begin{center}
   {\includegraphics[width=0.5\textwidth]{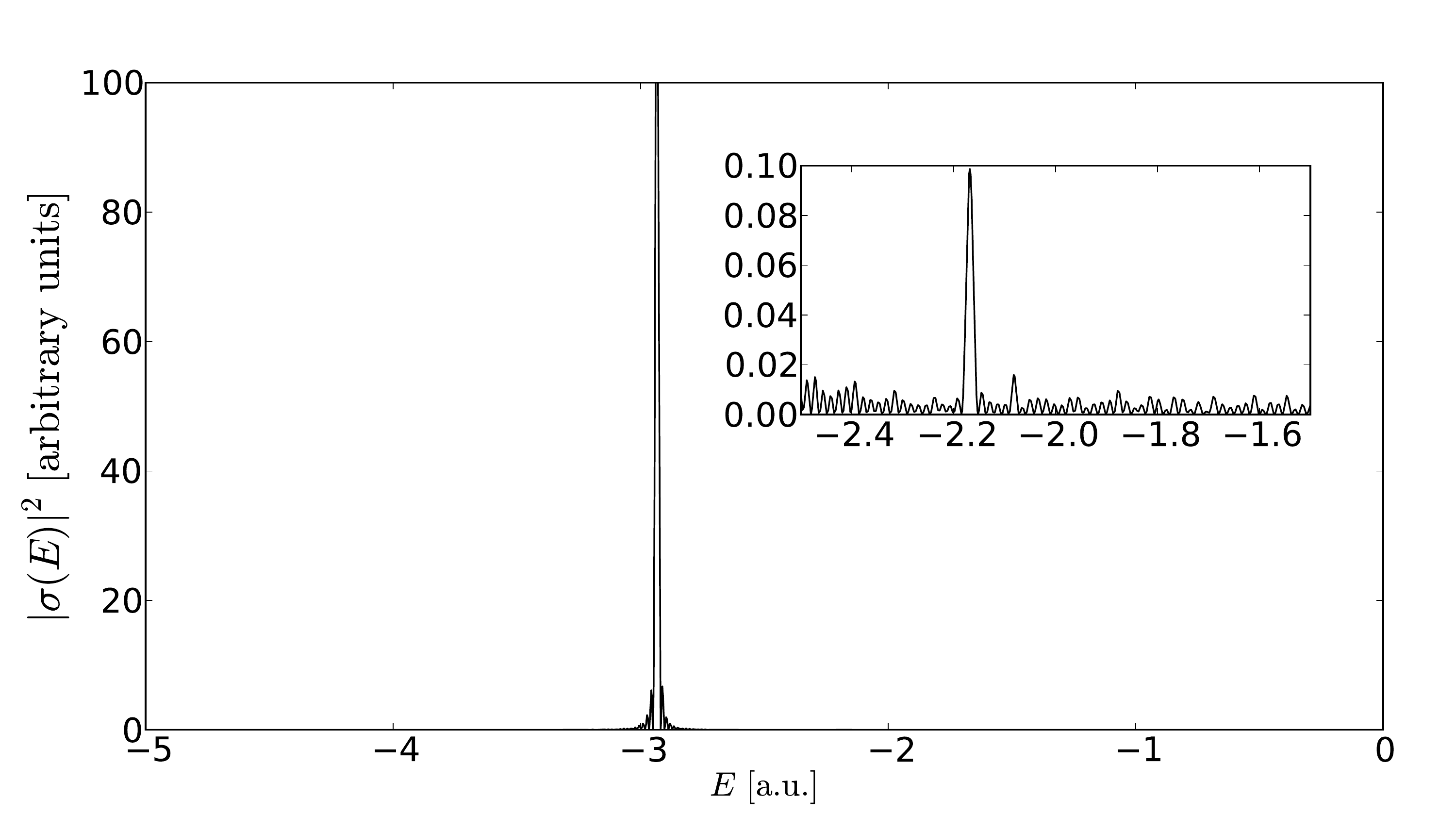}}
 \end{center}
  \caption[NumTestHeGSSpec]{Spectrum $\sigma(E)$ of Eq. \eqref{eq:HamFull} obtained through numerical integration of the TDSE for $t=200\,\mathrm{a.u.}$ with initial wave packet of Eq. \eqref{eq:PsiInitFull} and the grid parameters according to Eq. \eqref{eq:NumParam}.}
  \label{fig:NumTestHeGSSpec}
\end{figure}

Finally, we calculate the spectrum of the two-electron hamiltonian of Eq. \eqref{eq:HamFull} with all three degrees of freedom $(r_1,r_2,\theta)$.
Propagation of the initial wavepacket
\begin{equation}
 \psi(r_1,r_2,\theta,0) = r_1r_2e^{-r_1^2-r_2^2-(\theta-\pi)^2}
\label{eq:PsiInitFull}
\end{equation}
for a time of $t=200\,\mathrm{a.u.}$ yields the spectrum shown in Fig.\ref{fig:NumTestHeGSSpec}.

A distinct peak is visible at an energy of $E = -2.935\,\mathrm{a.u.}$ close to the accurate value for the ground state energy of the helium atom, $E_0 = -2.90372\,\mathrm{a.u.}$~\cite{pe62}, with a relative error of $10^{-2}$.
In addition, a second peak with an energy of $E=-2.17\,\mathrm{a.u.}$ can be identified, which is within 1\% error consistent with the energy of the first excited state,  $E_{(1s2s)\,^1S}=-2.14\,\mathrm{a.u.}$~\cite{pe62}.
Further peaks of the spectrum are not visible, due to the fact that the initial state of Eq. \eqref{eq:PsiInitFull} has small overlap with the respective eigenstates.

Hence, the crucial spectral features of the helium atom, the ground state energy and the energy gap to the first excited state, are described with sufficient accuracy  to perform reliable scattering calculations with the method described. 
\section*{References}
\def\harvardurl#1{\newline\texttt{#1}}\def\harvardurl#1{\relax}


%
\end{document}